\documentclass[preprint,12pt]{elsarticle}

\usepackage{amssymb}

\journal{New Astronomy}

\begin{document}

\begin{frontmatter}



\title{Multiwavelength study of VHE emission from Markarian 501 using TACTIC observations during April-May, 2012}


\author[1]{P Chandra}
\ead{chandrap@barc.gov.in}
\author[1]{K K Singh}\author[1]{R C Rannot}\author[1]{K K Yadav}\author[1]{H Bhatt} \author[1]{A K Tickoo} \author[1]{B Ghosal} \author[1]{M Kothari} \author[1]{K K Gaur} \author[1]{A Goyal} \author[1]{H C Goyal}  \author[1]{N Kumar}  \author[1]{P Marandi}  \author[1]{N Chouhan}  \author[1]{S Sahayanathan}  \author[1]{K Chanchalani}
 \author[1]{N K Agarwal} \author[1]{V K Dhar}  \author[1]{S R Kaul}  \author[1]{M K Koul}  \author[1]{R Koul} \author[1]{K Venugopal} \author[1]{C K Bhat} \author[1]{C Borwankar}
 \author[3]{J Bhagwan}
   \author[2]{A C Gupta}
\address[1]{Astrophysical Sciences Division, Bhabha Atomic Research Centre, \\
Mumbai - 400 085, India.}
\address[2]{Aryabhatta Research Institute of Observational Sciences.\\
Nainital- 263129, India.}
\address[3]{School of Studies in Physics and Astrophysics, Pt. Ravishankar Shukla\\
University, Amanaka G.E. Road, Raipur – 492010, India.}

\begin{abstract}
We have observed Markarian 501 in Very High Energy (VHE) gamma-ray wavelength band for 70.6 hours from 15 April  to 30 May, 2012 using TACTIC telescope. Detailed analysis of $\sim$66.3 hours of clean data revealed   the presence of a TeV $\gamma$-ray signal (686$\pm$77 $\gamma$-ray events) from the source direction with a statistical significance of 8.89$\sigma$ above 850 GeV. Further, a total of 375 $\pm$ 47   $\gamma$-ray like events were detected  in 25.2 hours of observation from 22 - 27 May, 2012 with a statistical significance of 8.05$\sigma$  indicating  
that the source has possibly switched over  to  a  relatively high gamma-ray emission state. We have derived  time-averaged differential energy spectrum of  the state   in the energy range 850 GeV - 17.24 TeV  which fits well with a  power law function of the form $dF/dE=f_{0}E^{-\Gamma}$ with $f_{0}=  (2.27 \pm 0.38) \times 10 ^{-11} $ photons cm$^{-2}$ s$^{-1}$ TeV$^{-1}$ and $\Gamma=2.57  \pm  0.15$. In order to investigate the  source state,  we have also used  almost  simultaneous multiwavelength   observations viz: high energy data collected by $\it{Fermi}$-LAT,  X-ray data collected by $\it{Swift}$-XRT and MAXI, optical and UV data collected by $\it{Swift}$-UVOT,  and radio data collected by OVRO, and reconstructed broad-band Spectral Energy Distribution (SED). The obtained SED supports   leptonic  model (homogeneous single zone) for VHE gamma-ray emission involving synchrotron and synchrotron self Compton (SSC) processes.
\end{abstract}


\end{frontmatter}



\section{Introduction} \label{introduction}
Blazars, a class of active galactic nuclei (AGN), are well known variable sources of radiation and currently represent a  dominant 
class of the extragalactic gamma-ray sky. These objects are further classified as BL Lacertae (BL Lac) and Flat Spectrum Radio Quasar (FSRQ)
with the mere difference of the presence of strong optical emission lines in the later. These objects are believed to be powered by accretion onto 
super massive black holes and have relativistic jets which are connected to the accretion disk and aligned very close to the observer's
line of sight \cite{Ury+95}. The broad-band emission from radio to $\gamma$-rays from these objects is dominated by strongly
Doppler-boosted non thermal radiation which is produced in the jets consisting of magnetised plasma.
The observed broad-band SED of blazars consists of two peaks with  first peak usually in the IR to X-ray,
while second one at $\gamma$-ray energies when plotted in the $\nu F_{\nu}$ versus $\nu$ representation   \cite{sam+96, fos+98}.
The non-thermal emission in the first peak is relatively well understood and ascribed as due to synchrotron radiation of relativistic electrons accelerated in the jets,
while the origin of the second peak is still unclear. In the leptonic scenario,  the high energy SED peak is attributed to Inverse Compton (IC)
scattering of either the synchrotron photons themselves called Synchrotron Self-Compton (SSC) emission or external photons called
External Compton (EC) emission or both by the same population of relativistic
electrons   \cite{ghi+98,der+92,sik+94}. 
Alternatively, the  hadronic models  explain the high energy peak of the blazar emission  
via pion decay with subsequent synchrotron and/or
IC emission from the decay products, or by synchrotron radiation from ultra-relativistic protons   \cite{mann+93,aho+00,pohl+00,mucke+01,mucke+03}.

Mrk 501 (z = 0.034) is a high-energy peaked blazar (HBL) discovered first in the energy range above 500 GeV by Whipple 
collaboration  \cite{Quinn+96}. Several aspects of the Very High Energy (VHE) $\gamma$-ray emission from  this blazar including 
the correlations between  intensity of X-rays, optical and TeV gamma-rays  have been studied \cite{Catanese+97}. 
These observations have reflected a variable nature of  the source in terms of both flux as well as spectrum, with a time scale as short as few minutes in some flares 
  \cite{abdo+11+501}. 
The variability time scales and flaring timescales can be the most direct way to probe the dynamics operating in
jet plasma, in particular compact regions of shock acceleration   \cite{Kataoka+01}.

Mrk 501 exhibited a strong temporal  and spectral variability in $\gamma$-ray flux during 1997 when it underwent  through a major VHE $\gamma$-ray flare at 10 times of the Crab nebula flux above 1TeV 
  \cite{dja+99,aha+99,amen+00}.
A fastest $\gamma$-ray flux variability on a timescale of
a few minutes was detected from it  in 2005 by Whipple collaboration  \cite{alb+07}.  
The source spectra have been  found to be harder at
brighter states as  compared to those during the source low-activity states   \cite{alb+07, godam+08, and+09}.
In May 2009, VERITAS collaboration  detected  a variable TeV $\gamma$-ray  emission at the flux level exceeding 
2 Crab    \cite{hua+09} from this source.
In  October 2011, Mrk 501 underwent a strong flare detected
by the MAXI satellite in X-rays   \cite{soo+11} and
by the ARGO-YBJ detector in VHE $\gamma$-rays   \cite{Bartoli-2011}. During a recent simultaneous radio-to-TeV observing campaign, Mrk 501 was 
observed both in quiescent and elevated state wherein the observed broad band states were reproduced by SSC model through a decrease in magnetic
field and increased luminosity and hardness of relativistic particles  \cite{Furniss-2015}.
H.E.S.S. observations of this source during 2012 and 2014 show rapid variability down to a time scale of four minutes in the 2-20 TeV energy range,
particularly 2014 observations have reported a flux level comparable to the 1997 historical flaring state  \cite{Colonga-2015}.
 A harder-when-brighter behaviour has been again reported in these observations.

In this paper, we present results of TeV $\gamma$-ray observations conducted during 
2012 using TACTIC $\gamma$-ray telescope along with results in other wavelength bands. In Section~\ref{tactic}, we briefly describe the TACTIC telescope and provide the details of Mrk 501 observations, data analysis procedure and results.
In Section~\ref{mwl}, we discuss multiwavelenght studies of the source including light curves, variability and spectral energy distribution during a possible high state period. Finally in Section~\ref{discuss}, we present discussion and conclusions. 

\section{TACTIC telescope} \label{tactic}
TACTIC (TeV Atmospheric Cherenkov Telescope with Imaging Camera) $\gamma$-ray telescope is located at Mt Abu 
(24.6$^\circ$N, 72.7$^\circ$E, 1300m asl) Rajasthan, India. It deploys a F/1 type tessellated parabolic tracking light collector
of $\sim$9.5 m$^2$ area consisting of 34 front-coated spherical glass facets of 0.6m diameter. The mirror facets have been pre-aligned to
produce an on-axis spot of $\sim$0.3$^\circ$ diameter at the focal plane. The focal  plane instrumentation of the telescope is a 349 pixel
(ETL 9083UVB) Photo-Multiplier Tube (PMT) based imaging camera which has an uniform pixel resolution of $\sim$0.3$^\circ$.  The camera records images of the atmospheric Cherenkov showers with a total field of view of $\sim$6$^\circ$ $\times$ 6$^\circ$.

Data used in this work have been collected with the inner 225 pixels and the innermost 121 pixels were
used for generating the event trigger. The trigger is based on the three nearest neighbour triplets logic with single pixel
threshold set to 14 photo-electrons (pe) \cite{aktickoo2014}. The triggered events are digitized by CAMAC based 12-bit charge to digital converters (CDC) which have a full scale range of 600pC. The safe anode current (3$\mu$A) operation of the PMT has been ensured by implementing a gain control algorithm \cite{BhattN2001}. 
The detailed description of the telescope related hardware and software can be found in \cite{yadavkk2004,rkoul2007}.

Major upgradation work was carried out for TACTIC during the fall of 2011 involving replacement of high voltage and signal cables and installing new compound parabolic concentrators (CPC) on the imaging camera of the telescope in order to increase the photon collection efficiency.  
These CPCs have square entry and circular exit aperture with $\sim$85\% reflectivity in the wavelength range 400 - 500 nm. 
A dedicated CCD camera was also installed for detailed pointing runs which gives source position with an accuracy of $\sim$3 arc-min on the camera plane. The telescope is sensitive to $\gamma$-rays above 850 GeV and can detect the Crab Nebula at 5$\sigma$ significance level in 12 hours of observation. More detail about the up-gradation of TACTIC can be found in     \cite{aktickoo2014}.

\subsection{TeV Observations of Mrk 501  and data analysis} \label{mrk501observationtactic}
TeV observations on Mrk 501 were taken during 15 April - 30 May, 2012. Total duration of observations on the source during above mentioned 
observation period was $\sim$70.6 hours. Several standard data quality checks have been used to evaluate the overall system behavior and the general quality of the recorded 
data. These include conformity of the prompt coincidence rates with the expected zenith angle dependence, compatibility of the arrival times of prompt coincidence events 
with Poissonian statistics and the behavior of the chance coincidence rate with time.
After applying these data quality checks, we have selected good quality data sets of 66.3 hours as reported in detail in Table~\ref{tab:details_mrk501}.  
\begin{table}[h]
\begin{center}
\caption{Details of Mrk 501 observations using TACTIC during April-May 2012.}
  \begin{tabular}{ c c c c  }
    \hline
     Month, Year & Observation Dates & Total Data & Data Selected  \\
      &  & (hours) & (hours)  \\
    \hline
            April, 2012               &         15,16,20,22,23,25,26          &     13.4            &        11.6               \\
            May, 2012                 &         11,14-20,22-30                &     57.2            &        54.7               \\    
  \hline
            April-May, 2012            &                                      &       70.6          &        66.3 \\
  \hline
      \end{tabular}%
  \label{tab:details_mrk501}
  \end{center}
\end{table}
Detailed analysis of an imaging atmospheric Cherenkov telescope data, involves a number of steps including image cleaning, 
accounting for the differences in the relative gains of the PMTs, image parameterization, event classification and energy determination. 
In the first step, the CDC pedestals are subtracted from the CDC counts of each pixel of the camera. 
In order to select pixels containing Cherenkov image and to reduce noise contribution, we have used picture threshold (PT) of 6.5$\sigma$ and boundary threshold (BT) of 2.9$\sigma$. 
Here $\sigma$ is standard deviation of CDC distribution derived using 2000 events recorded for night sky background.
In the second step, relative gain calibration is performed by recording 2000 events from a pulsed light source (LED) in front of the camera surface during the data taking.  
The calibration data collected are then used to determine the relative gains of each pixel by comparing their mean signals with respect to a reference pixel. 
The next step in the data analysis is the image parameterization and event selection. Characterization of the Cherenkov images 
formed at the focal plane is performed using the moment analysis    \cite{hillas1985,hillas1998} and various image parameters, viz., Length (L),
Width (W), Distance (D), Frac2 (F2), Asymmetry, Alpha ($\alpha$), Size (S) etc     \cite{weekes1989} are calculated. The images from gamma-ray initiated showers are roughly elliptical in shape and described by the L and W parameters and its location and orientation within the telescope field of view are given by the D and $\alpha$ parameters, respectively.
\begin{table}[ht]
\caption{Dynamic supercuts selection criteria used in the present work.}
\begin{center}
\begin{tabular}{|c|c|}
\hline
 BT=2.9$\sigma$  and  PT=6.5$\sigma$ \\ \hline
 1pe $\approx$ 6.2 CDC Counts \\ \hline
 S$>$50pe \\ \hline
 0.53 $\times$ $Cos^{0.47}\theta$  $\leq $ D $\leq $ 1.14 $\times$ $Cos^{-0.41}\theta$; $\theta$=Zenith Angle\\ \hline
 Minimum number of pixels $\geq 4$ \\ \hline
 F2 $>$ 0.29 + 0.1146 $\times\theta$  \\ \hline
 0.16$\leq$L$\leq$[0.1446+0.07179$\times\theta$]+[0.0683-0.0375$\times\theta\times$log(S)]\\\hline
 0.04$\leq$W$\leq$[-0.1436+0.3217$\times\theta$]+[0.0738-0.0790$\times\theta\times$log(S)]\\\hline
 L/W $>$ 1.55, Asymmetry  $>$ 0.0 \\ \hline    
\end{tabular}
\end{center}
\label{tab:cuts}
\end{table} 

The standard dynamic supercut procedure is then used to separate $\gamma$-ray-like images from the overwhelming 
background of cosmic rays.  This procedure uses the image shape parameters L and W as a function of the image size and zenith angle so that energy
and zenith angle dependence of these parameters can be taken into account. The $\gamma$-ray selection criteria used in this analysis are given in Table~\ref{tab:cuts}.
The shape selected images are used to generate $\alpha$-distribution with its gamma-domain range of $\leq$ 18$^{0}$ for the TACTIC telescope. 
The contribution of the background events is estimated from a reasonably flat $\alpha$ region of 27$^{0}$ $\le$ $\alpha$ $\le$ 81$^{0}$. The number of $\gamma$-ray events is then calculated 
by subtracting the expected number of background events, calculated on the basis of the background region, 
from the $\gamma$-ray domain events. 
The significance of the excess events is calculated using the maximum-likelihood-ratio method of Li and Ma    \cite{lima1983}.

\subsection{TACTIC results} \label{tacticresult}

\begin{table}[ht]
\caption{Monthly spell wise analysis of Mrk 501 data recorded during April - May, 2012.}
\begin{center}
\begin{tabular}{|c|c|c|c|c|}
  \hline
     Spell & Observation Period & Time & Excess events & Sig. \\
     & & (hours) &  & ($\sigma$) \\
    \hline
            1           &         15-26 April, 2012       &           11.6      &   110 $\pm$ 35                  &     3.19		\\
            2           &         11-30 May, 2012         &           54.7      &   576 $\pm$ 69                  &     8.35          \\    
  \hline
                        &                                 &           66.3      &   686 $\pm$ 77                  &     8.89    \\
  \hline
\end{tabular}
\end{center}
\label{tab:analysis_mrk501_spell}
\end{table}
The data collected on Mrk 501 during 15 April - 30 May, 2012 were divided into two monthly spells and analysed using the analysis procedure mentioned above. 
A total of 686 $\pm$ 77 $\gamma$-ray like events were detected with statistical significance of 8.89$\sigma$. 
As evident from Table~\ref{tab:analysis_mrk501_spell}, the most of the $\gamma$-ray like events (576 $\pm$ 69) were detected during the 
spell 2 with statistical significance of 8.35$\sigma$. 

\begin{figure}[h]
    \centering
   \includegraphics[width=1.0\textwidth]{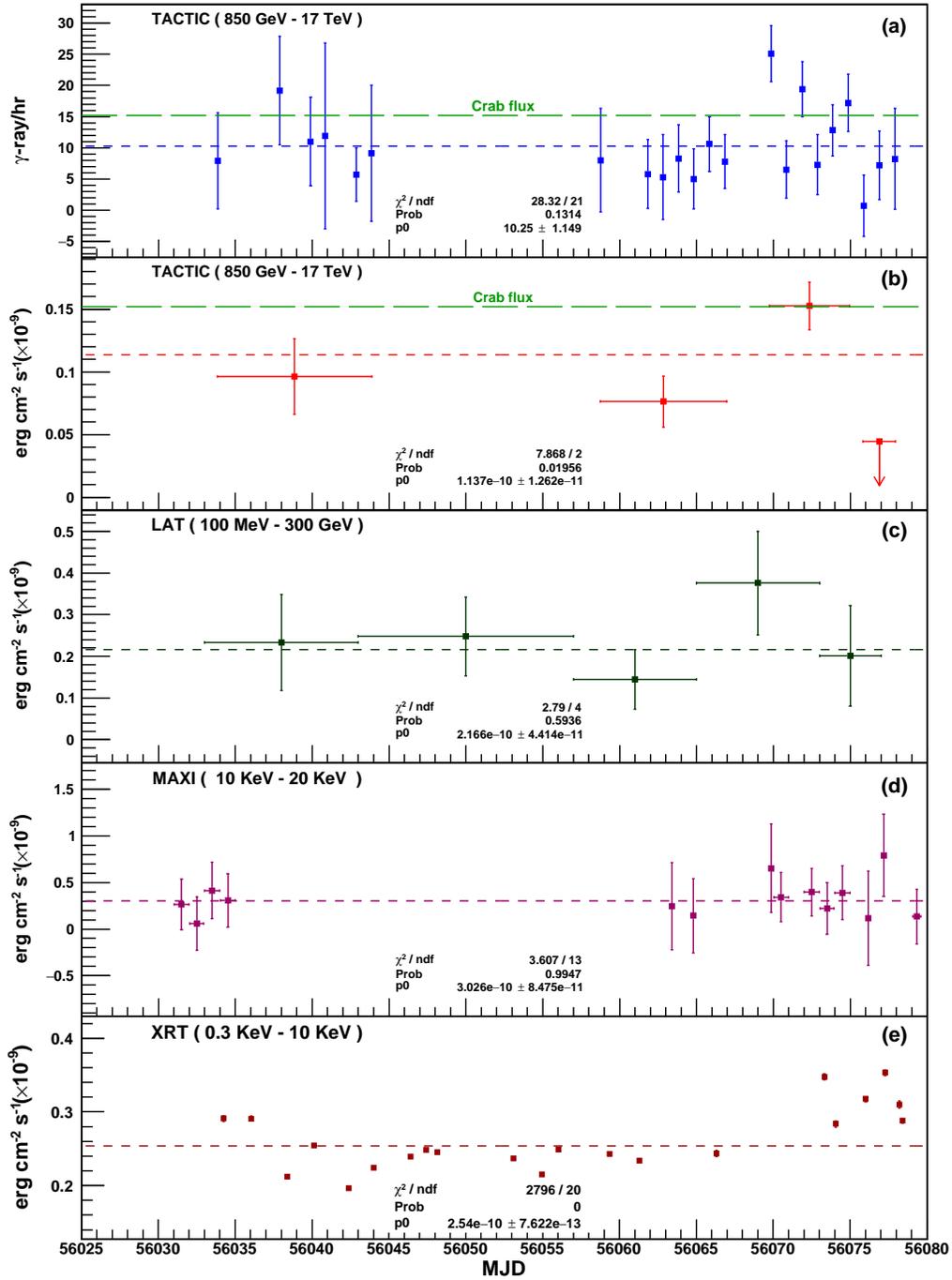}
    \caption{$\gamma$-ray and X-ray light curves of Mrk 501 during April - May, 2012. The dotted line represent the average emission level
from the source in different bands.}
    \label{fig:lightcurve_mrk501}
\end{figure}

Further, daily TACTIC light curve as shown in Figure~\ref{fig:lightcurve_mrk501}(a) indicates, the source flux levels were more than 1 Crab Unit (CU) on few days during 22 - 27 May, 2012.
This relatively high emission state of Mrk 501 is also clearly visible from Figure~\ref{fig:lightcurve_mrk501}(b). 
\begin{table}[ht]
\caption{Details of Mrk 501 data analysis results of TACTIC observations during 22 - 27 May, 2012.}
  \begin{tabular}{ c c c c c c}
    \hline
     \multicolumn{1}{p{2.1cm}}{\centering Obs. date \\ (May, 2012)} & Start MJD & End MJD & $\gamma$-rays  & Sig.($\sigma$) &  \multicolumn{1}{p{2.1cm}}{\centering Obs.  \\ Time (hours)} \\
     
    \hline
	        22  &   56069.75	&	56069.95            &    119 $\pm$  22        &   5.54      &	4.7	\\
	        23  &   56070.74	&	56070.95            &     29 $\pm$  20        &   1.42      &	4.5	\\            
	        24  &   56071.77	&	56072.00            &     84 $\pm$  19        &   4.44      &	4.3	\\            
	        25  &   56072.77	&	56072.95            &     30 $\pm$  20        &   1.53      &	4.1	\\            
	        26  &   56073.77	&	56073.96            &     52 $\pm$  17        &   3.13      &	4.1	\\            
	        27  &   56074.78	&	56074.95            &     61 $\pm$  16        &   3.74      &	3.5	\\            
	           \hline         
		&  56069.75	&	56074.95            &     375 $\pm$ 47        &   8.05      &	25.2	\\            
  \hline         
   \end{tabular}
   \label{tab:analysis_mrk501_may2012}
\end{table} 

\begin{figure}[h]
    \centering
   \includegraphics[width=1.0\textwidth]{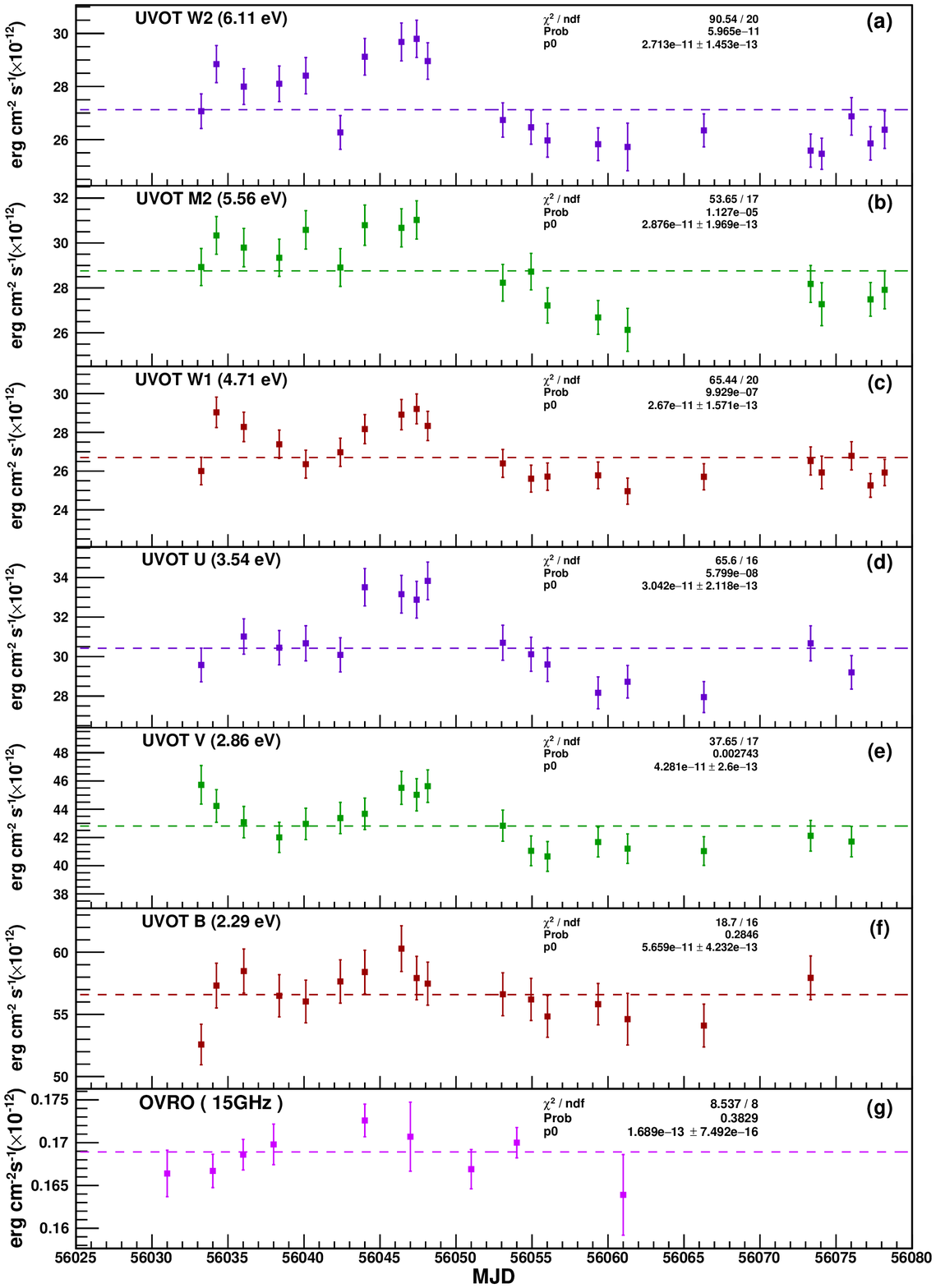}
    \caption{Light curves of Mrk 501 in UV, optical and radio energy bands during April-May, 2012. The dotted line represent the average emission level
from the source in different bands.}
    \label{fig:lightcurve_mrk501_2}
\end{figure}

\begin{figure}[h]
    \centering
    \includegraphics[width=0.8\textwidth]{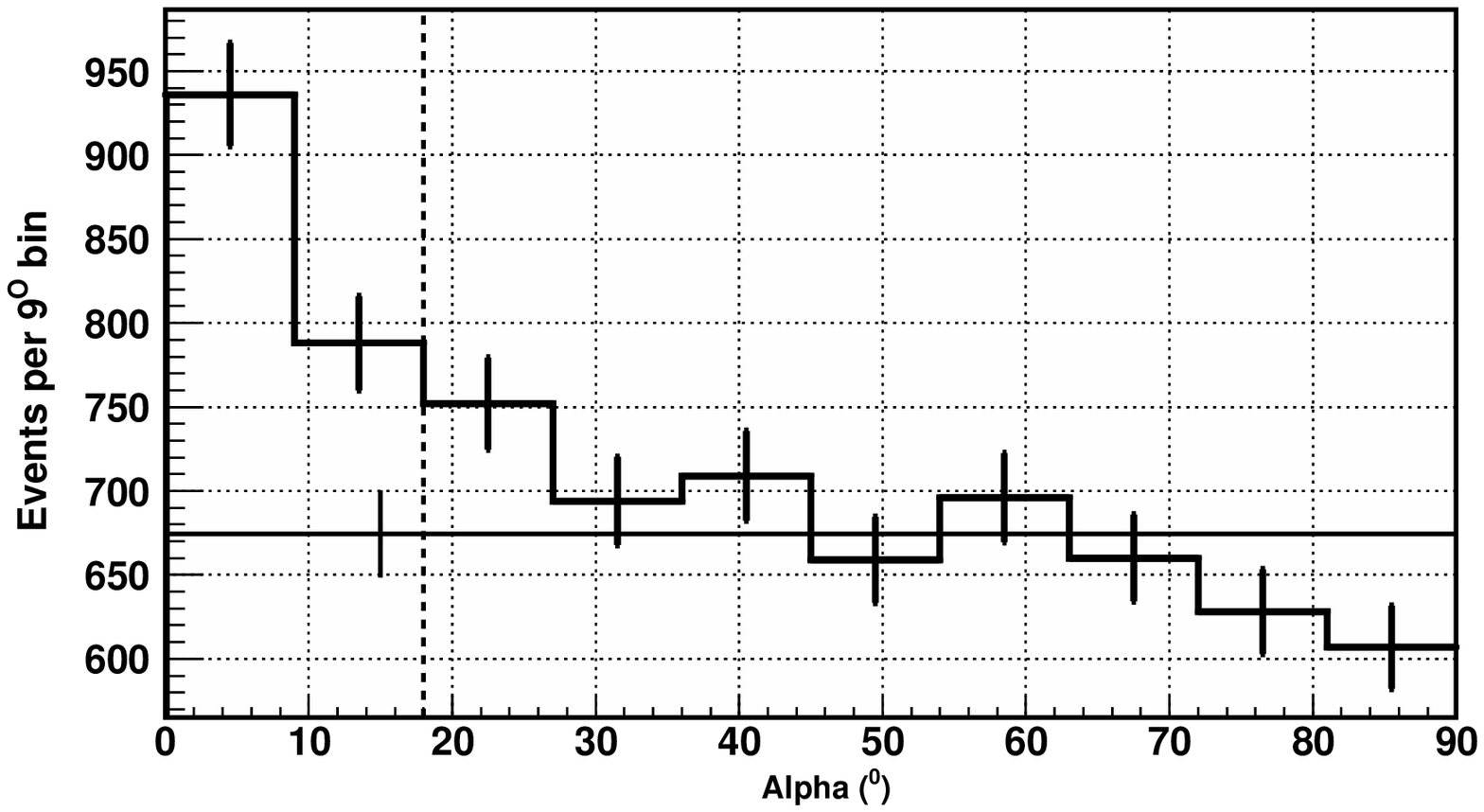}
    \caption{    Distribution of the image parameter $\alpha$ for the TACTIC observations of Mrk 501 during
22 - 27 May, 2012. The vertical dashed line represents the $\alpha$ cut used for estimating the number of events in $\gamma$-ray domain.
The horizontal line represents the mean background level per 9$^{0}$ bin derived using the reasonably
flat $\alpha$ region of 27$^{0}$ $\le$ $\alpha$ $\le$ 81$^{0}$ . Error bars shown are statistical errors only.}
    \label{fig:alpha_mrk501}
\end{figure}

During this period a total of 375 $\pm$ 47  $\gamma$-ray like events were detected with statistical significance of 8.05$\sigma$ in 25.2 
hours of observation. 
During this source state, an average hourly $\gamma$-ray rate was 14.88 $\pm$ 1.85 which is around 0.98 CU 
(where 1 CU $\simeq$ 15.2 $\pm$ 1.2 $\gamma$-rays/hour for TACTIC) as shown in Table~\ref{tab:table_lightcurve_avg_mrk501}.  
Therefore for further analysis, we have considered the data collected during 22 - 27 May, 2012. 
The detailed analysis of this period data has been given in Table~\ref{tab:analysis_mrk501_may2012} and distribution of $\alpha$ parameter has
been shown in Figure~\ref{fig:alpha_mrk501} after applying shape and orientation related imaging cuts as given in Table~\ref{tab:cuts}.
\begin{table}[h]
\caption{Details of Mrk 501 divided into four spell using TACTIC during period April-May 2012.}
  \begin{tabular}{c c c c c c r r}
    \hline
    Spell& Start  & End & Time & Dates  & $\gamma$-rays & $\gamma$-rays/hr   & Sig. \\  
      No.&  MJD   & MJD &   (hours)     &               &            &        &     ($\sigma$)   \\
    \hline
            1 & 56033 & 56043 & 11.6 & 16-26 April 2012 & 110 $\pm$ 35 &  9.48 $\pm$ 3.02   & 3.19    \\            
            2 & 56058 & 56066 & 23.4 & 11-19 May 2012 & 175 $\pm$ 47   &  7.48 $\pm$ 2.01  & 3.74    \\
            3 & 56069 & 56074 & 25.2 & 22-27 May 2012 & 375 $\pm$ 47   & 14.88 $\pm$ 1.87  & 8.05    \\
            4 & 56075 & 56077 &  5.9 & 28-30 May 2012 &  26 $\pm$ 20   &  4.41 $\pm$ 3.39  & 1.29    \\
  \hline
      \end{tabular}%
     \label{tab:table_lightcurve_avg_mrk501}
\end{table}

\subsection{VHE $\gamma$-ray differential energy spectrum} \label{differntialspectrum}
We have estimated the time-average differential energy spectrum of Mrk 501 using TACTIC observations conducted during the period 22- 27 May, 2012. The energy reconstruction procedure uses artificial neural network with
3:30:1 configuration and resilient backpropagation algorithm. The energy of each $\gamma$-ray-like
event is estimated on the basis of its image size, distance and zenith angle. This method yields
an energy resolution of $\sim$26$\%$, the details of which have been given in ~\cite{dhar2009,yadav2007}. 

\begin{figure}
    \centering
    \includegraphics[width=1.0\textwidth]{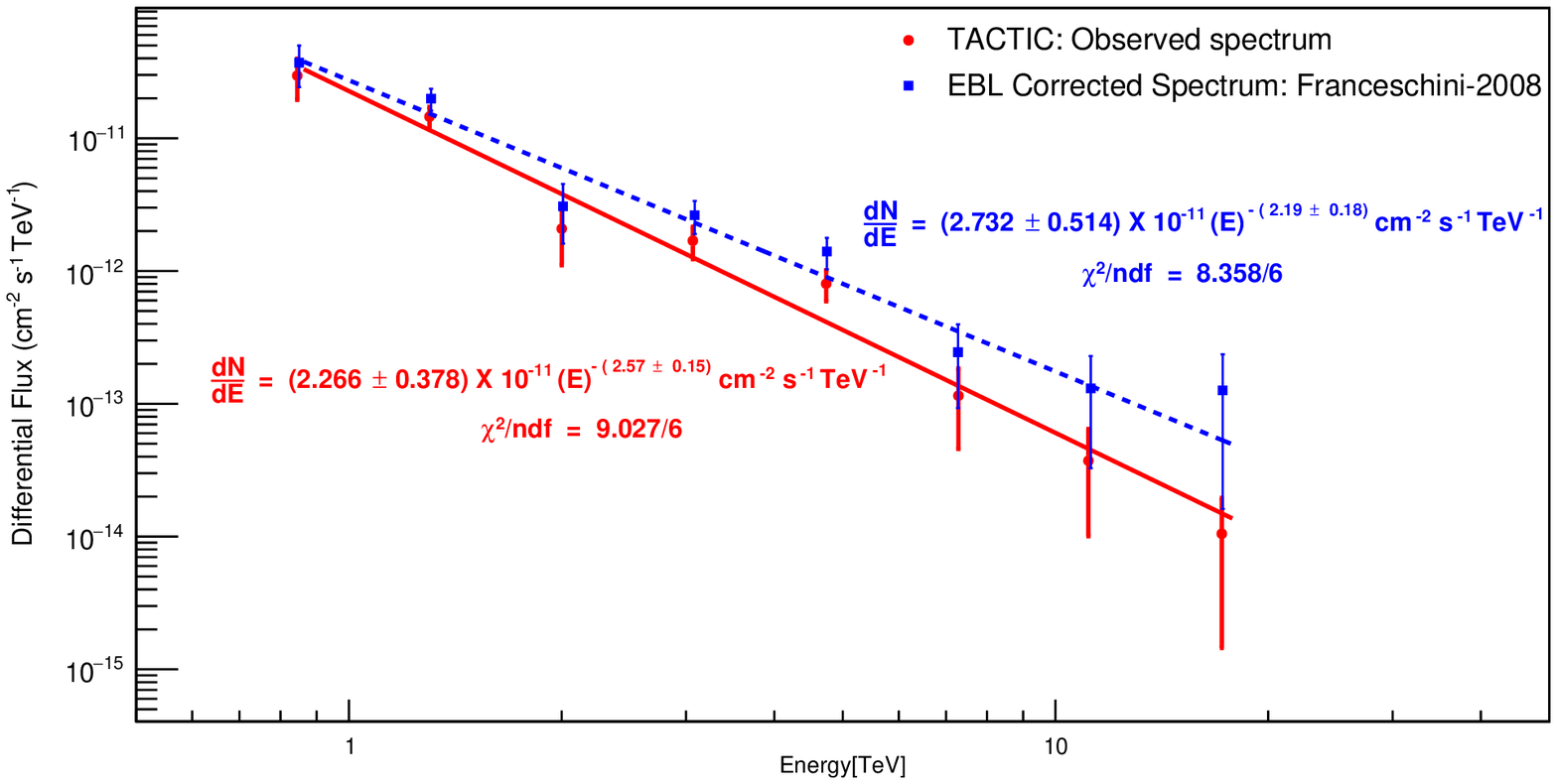}
    \caption{Differential energy spectrum of Mrk 501 during high state from 22 to 27 May, 2012. The observed differential energy spectrum is fitted with power-law and shown in red color. EBL absorption of TeV $\gamma$-ray photons has been estimated using Franceschini model ~\cite{franceschini2008} and fitted with power-law function and parameters are given on the figure in blue color.} 
        \label{fig:spectrum_mrk501_high}
\end{figure}

The measured differential energy spectrum calculated for 22 - 27 May, 2012 (MJD 56069-56074) in energy range 850 GeV - 17.24 TeV is compatible with a power law fit of the form $dF/dE=f_{0}E^{-\Gamma}$ with $f_{0}=  (2.27 \pm 0.38) \times 10 ^{-11} $ ph cm$^{-2}$ s$^{-1}$ TeV$^{-1}$ with $\Gamma=2.57  \pm 0.15$ ($\chi^2$/ndf = 9.027/6; probability=0.172). 
In order to obtain the intrinsic $\gamma$-ray source spectrum, 
the absorption of TeV $\gamma$-ray photons from Mrk 501 due to EBL has been estimated using the
Franceschini model~\cite{franceschini2008}. The EBL corrected spectrum is further described by power law as shown in Figure~\ref{fig:spectrum_mrk501_high} for the high state of the source.
\begin{table}
\caption{Differential energy spectrum data for Mrk 501 during 22-27 May, 2012 with the TACTIC telescope. }
    \centering
  \begin{tabular}{ c c c }
    \hline
    Energy & Differential flux  & EBL corrected flux \\  
     (TeV) & (photon cm$^{-2}$ s$^{-1}$ TeV$^{-1}$) & (photon cm$^{-2}$ s$^{-1}$ TeV$^{-1}$)\\
    \hline
	  0.85    &  (2.92 $\pm$ 1.01)$\times10^{-{11}}$      & (3.71 $\pm$ 1.28)$\times10^{-{11}}$\\
	  1.31    &  (1.43 $\pm$ 0.27)$\times10^{-{11}}$      & (1.99 $\pm$ 0.37)$\times10^{-{11}}$\\
	  2.01    &  (2.05 $\pm$ 0.98)$\times10^{-{12}}$      & (3.07 $\pm$ 1.46)$\times10^{-{12}}$\\
	  3.09    &  (1.65 $\pm$ 0.46)$\times10^{-{12}}$      & (2.62 $\pm$ 0.73)$\times10^{-{12}}$\\
	  4.75    &  (7.91 $\pm$ 2.11)$\times10^{-{13}}$      & (1.40 $\pm$ 0.37)$\times10^{-{12}}$\\
	  7.30    &  (1.11 $\pm$ 0.69)$\times10^{-{13}}$      & (2.42 $\pm$ 1.52)$\times10^{-{13}}$\\
	 11.22   &   (3.63 $\pm$ 2.72)$\times10^{-{14}}$      & (1.31 $\pm$ 0.98)$\times10^{-{13}}$\\
	  17.24  &   (1.04 $\pm$ 0.91)$\times10^{-{14}}$      & (1.26 $\pm$ 1.10)$\times10^{-{13}}$\\
	 \hline
      \end{tabular}%
  \label{tab:mrk501_spectrum_points_22-27May2012}
  \label{tab:addlabel}%
\end{table}

\section{Multi-wavelength observations of Mrk 501} \label{mwl}
The multi-wavelength data for Mrk 501 have been obtained from several instruments in high energy $\gamma$--rays, X-rays, 
optical and radio energy bands thanks to the \emph{Fermi} multi-wavelength blazar monitoring program\footnote{{http://fermi.gsfc.nasa.gov/ssc/observations/multi/programs}}.

\subsection{High Energy (HE) $\gamma$--ray data}
The HE $\gamma$--ray data for photon energies greater than 20 MeV are provided by Large Area Telescope (LAT) onboard 
\emph{Fermi} satellite    \cite{Atwood2009}. The \emph{Fermi}-LAT is a pair conversion detector surveys the 
whole sky every three hours with a field of view of 2.4 sr and thus provides the most sensitive and complete coverage of 
$\gamma$--ray sky in its nominal scanning mode. LAT data used in this work are publicly available reprocessed Pass 8 data 
downloaded during April 15, 2012 (MJD 56032) to May 27, 2012 (MJD 56074)\footnote{{http://fermi.gsfc.nasa.gov/ssc/data}}. We have performed an unbinned maximum likelihood analysis
to obtain the light curve in the energy range 0.1-300 GeV with the standard analysis tool \emph{gtlike}, implemented in the 
Science Tools software package (version v10rop5). A refined version of the instrument response functions P8R2$\_$SOURCE$\_$V6
with improved understanding of the point-spread function and effective area are used in the analysis. We have selected events/photons 
in a circular region of interest with 15$^\circ$ radius centered at the position of Mrk 501 from the third \emph{Fermi} gamma-ray 
source (3FGL) catalog \cite{Acero2015}. In addition, we have excluded events arriving from zenith angles 
above 90$^\circ$ to reduce the contamination from $\gamma$--ray bright Earth limb and time intervals when the Earth entered 
the LAT field of view selecting photons with spacecraft rocking angle less than 52$^\circ$. To model the Galactic and isotropic 
diffuse background (sum of residual cosmic-ray background and extragalactic diffuse $\gamma$--ray background), we have used the 
files \emph{gll$\_$iim$\_$v06.fits} and \emph{iso$\_$P8R2$\_$SOURCE$\_$V6$\_$v06.txt} provided with \emph{Fermi} Science Tools. All point sources in 
the 3FGL catalog within 20$^\circ$ of Mrk 501 (3FGLJ1653.9+3945), including the source itself, have been considered in the likelihood 
analysis. For obtaining the light curves, sources within the region of interest (15$^\circ$) are fitted with the power law models keeping 
spectral indices and normalization as free parameters whereas parameters of the sources beyond the radius of region of interest are fixed 
to the values given in the 3FGL catalog. 
\par
The spectral analysis of Mrk 501  is performed in the energy band 0.1 - 300 GeV for the time periods 22-27 May, 2012 (MJD 56069-56074). 
A power law model fitted to the data between minimum and maximum energies of the bin with the normalization as free parameter and spectral indices set to the value obtained from the integration of data over the above time period in the energy range 0.1 - 300 GeV. 
The time-averaged differential energy spectrum of the source in this energy band fits well with the power law function of photon spectral index 1.65 $\pm$ 0.21.
\subsection{X-ray data}
The near simultaneous data from X-ray observations of Mrk 501 during the above mentioned period have been collected from the telescopes onboard \emph{Swift} 
and \emph{MAXI} satellites. The X-Ray Telescope (XRT)       \cite{Burrows2005} onboard the \emph{Swift} satellite provides observation
of soft X-rays in 0.3-10 keV energy range. We have processed the XRT data following the standard \emph{FTOOLS} distributed by HEASARC within the 
HEASoft package (v6.19). We have produced cleaned event files from the observations in window timing (WT) mode with the standard filtering criteria using the 
task \emph{xrtpipeline} (ver 0.13.2)  with latest available calibration files  for \emph{Swift} (ver 20160121) . The source photons have been extracted 
from a circular region of radius 45$^{\prime\prime}$ and background with similar radius is selected from the nearby source free region. The 
ancillary response files (ARFs) correcting for exposure map are generated with the \emph{xrtmkarf} tool. Finally, we have extracted the spectra in the 
energy range 0.3-10 keV from the corresponding event files  and binned using \emph{grppha} to ensure a minimum of 30 counts per bin for reliable use 
of $\chi^2$-statistic. The spectra are fitted with a power law model consisting of photo-electric absorption in the Milky Way due to neutral hydrogen 
using \emph{xspec} (ver 12.9.0) model \emph{phabs$\ast$zpow} at redshift z=0.034. The neutral hydrogen column density ($N_H$) for line of sight  
absorption is fixed to the Galactic value 1.55$\times$10$^{20}$ cm$^{-2}$ using LAB model \cite{Kalberla2005}. The XRT light curves and spectra of the source have been obtained for daily observations following the above procedure. We have also used archival data in the energy range 10-20 keV from X-ray instrument onboard 
\emph{MAXI} satellite \cite{Matsuoka2009} from the website\footnote{{http://maxi.riken.jp/mxondem/}}. 

\subsection{Ultraviolet, Optical and Radio data}
Mrk 501 is observed at UV, optical and radio frequencies by different instruments world wide as part of \emph{Fermi} blazar monitoring 
program. In this work, we have analyzed all the observations performed by the \emph{Swift} Ultra Violet and Optical Telescope \cite{Roming2005} between MJD 56032 and 56078. The instrument cycled through each of the three ultraviolet pass band filters UVW1, UVM2 and UVW2 and three optical pass band filters V, B and U. We have used only level 2 image mode data in which image is directly accumulated onboard, discarding the photon timing information, and reduced the telemetry volume. The data have been processed with the standard procedure\footnote{http://www.swift.ac.uk/analysis/uvot/} using UVOTMAGHIST task of HEASoft package. The UVOT source fluxes are extracted from a circular region centered on the source position with 5$''$ radius, while the background is extracted from a nearby larger, source free, circular region of 10$''$ radius. The Mrk 501 fluxes corresponding to six UV/optical filters have been corrected for galactic reddening (E(B-V) = 0.019 mag; \cite{Schlegel1998}) and each spectral band is corrected for galactic extinction \cite{Roming2009}. The archival radio data at 15 GHz from the 40 meter telescope \cite{Richards2011} of the Owens Valley Radio Observatory 
(OVRO)\footnote{{http://www.astro.caltech.edu/ovroblazars/data}} have also been used for multi-wavelength study of the source.

\subsection{Light Curve (LC) analysis}
The multi-wavelength light curves of Mrk 501 from TeV $\gamma$--rays to radio are shown in the Figures~\ref{fig:lightcurve_mrk501} and~\ref{fig:lightcurve_mrk501_2} during the period April 15 - May 31, 2012 (MJD
56032-56078). For a clear representation, $\gamma$--ray and X-ray observations of the source have been reported in Figure~\ref{fig:lightcurve_mrk501}(a-e) while UV, optical and radio measurements are given in Figure~\ref{fig:lightcurve_mrk501_2}(a-g). Figure~\ref{fig:lightcurve_mrk501}(a) shows the daily TeV $\gamma$--ray rates detected with TACTIC telescope during the above period. Since the daily observations with TACTIC are not statistically significant, the entire TACTIC observations have been combined in four groups with different time intervals (Table~\ref{tab:table_lightcurve_avg_mrk501}) as shown in the Figure~\ref{fig:lightcurve_mrk501}(b). The first two flux points in the Figure~\ref{fig:lightcurve_mrk501}(b) correspond to the TACTIC observations with statistical significance above 3$\sigma$. The third flux point has been averaged over the period May 22-27, 2012 (MJD 56069-56074) and this corresponds to the statistical significance of 8.05$\sigma$. The last flux point in the light curve is not statistically significant and it corresponds to 99\% confidence level upper limit on the integral flux above 850 GeV from the source. The average VHE $\gamma$--ray emission from Mrk 501 during the period of TACTIC observations is characterized by a mean flux level of (1.14$\pm$0.13)$\times 10^{-10}$ erg cm$^{-2}$ s$^{-1}$ with $\chi^{2}$ value of 7.87 for 2 degrees of freedom (probability of this constant emission is 0.02). From the VHE $\gamma$--ray light curve in Figure~\ref{fig:lightcurve_mrk501}(b), the source was observed in a relatively high emission state during May 22-27, 2012.
The HE $\gamma$--ray emission in the energy range 100 MeV-300 GeV observed with \emph{Fermi}-LAT is shown in the Figure~\ref{fig:lightcurve_mrk501}(c). The flux points in the light curve are the time averaged observations over the time intervals overlapping with TACTIC flux points. All the five flux measurements reported in Figure~\ref{fig:lightcurve_mrk501}(c) correspond to the statistical significance above 5$\sigma$. A constant fit to the flux points gives an average flux (2.17$\pm$0.44)$\times 10^{-10}$ erg cm$^{-2}$ s$^{-1}$ with $\chi^{2}$ /ndf of 2.79/4 (probability of constant emission is 0.60). A relatively high emission state of the source during May 22-27, 2012 is also evident from the \emph{Fermi}-LAT light curve. The $\gamma$--ray observations of Mrk 501 with TACTIC and \emph{Fermi}-LAT have also been complemented by X-ray observations in two energy bands 0.3-10 keV and 10-20 keV using \emph{Swift}-XRT and MAXI respectively. The X-ray light curve observed with MAXI in the energy range 10-20 keV is shown in Figure 1(d). The X-ray emission from MAXI observations is observed to be consistent with a constant flux (3.03$\pm$0.85)$\times 10^{-10}$ erg cm$^{-2}$ s$^{-1}$ and $\chi^{2}$/ndf of 3.61/13 (probability of no variability is 0.99). The observations in the energy range 0.3-10 keV with \emph{Swift}-XRT are depicted in the Figure~\ref{fig:lightcurve_mrk501}(e). This light curve is observed to be highly variable and a constant fit to the flux points gives an average flux (2.54$\pm$0.08)$\times$10$^{-10}$  erg cm$^{-2}$ s$^{-1}$ with $\chi^2$/ndf of 2796/20 (probability of constant emission is almost zero).

Figure~\ref{fig:lightcurve_mrk501_2}(a-g) shows the daily light curves of Mrk 501 in UV, optical and radio wavelength bands for the period April 15, 2012 - May 31, 2012
(MJD 56032-56078). The results of constant fit to the observed flux points in different bands are indicated in the respective Figure~\ref{fig:lightcurve_mrk501_2}(a-g).
From the Figure~\ref{fig:lightcurve_mrk501_2}(a-c) it is evident that the UV emission from the source in W2, M2 and W1 bands is not compatible with constant fit (probability of constant flux is very low). Similar behaviour is also visible in optical U and V bands (Figure~\ref{fig:lightcurve_mrk501_2}(d-e)). However, the optical emission in B band is characterized by constant emission with $\chi^2$/ndf of 18.7/16 (corresponding probability of constant fit is 0.28). The radio observation
at 15 GHz by OVRO 40 meter telescope is also compatible with constant emission during the above time period. The radio flux points reported in
Figure~\ref{fig:lightcurve_mrk501_2}(g) are characterized by an average flux (1.69$\pm$0.07)$\times$10$^{-13}$  erg cm$^{-2}$ s$^{-1}$ and $\chi^2$/ndf value of 8.54/8
(probability of constant fit is 0.38).

It is interesting to compare the multiwavelength emission from Mrk 501 during April-May 2012 presented in this work with the recent observations
of the source using different instruments. Furniss et al. (2015) have reported near simultaneous multiwavelength observations of Mrk 501 between April 1, 2013
(MJD 56383) and August 10, 2013 (MJD 56514) \cite{Furniss2015}. The VHE $\gamma$--ray observations during this period with VERITAS and MAGIC are found to be
consistent with the range of emission states of Mrk 501 observed in past. The observations with MAGIC telescope hint for intra-night variability
during two nights and a constant fit to the whole light curve results in $\chi^2$/ndf of 57/4. The VERITAS observations of Mrk 501 in 2013 are
observed to vary by no more than a factor of three throughout entire observation. The HE observations with \emph{Fermi}-LAT during April-August 2013
are also found to be consistent with our findings in the current work. The daily light curves detected with \emph{Fermi}-LAT during the above period
are not statistically significant and therefore authors have reported weekly and 3.5-day binned light curves in their study \cite{Furniss2015}. The X-ray observations of Mrk 501 in the energy bands 0.3-10 keV and 3-79 keV by \emph{Swift}-XRT and \emph{NuSTAR} respectively during the above period have been investigated by Furniss et al. (2015) \cite{Furniss2015}. The authors have reported that the hard X-ray emission in the energy range 3-79 keV detected by \emph{NuSTAR} is variable on hour timescales and can be described by a log-parabolic spectrum. \emph{Swift}-UVOT observations of the source also indicate no variability in any filter during 2013 observations. The radio observations of Mrk 501 reported at different frequencies during the above time interval have also been found to be consistent with constant emission. The $\gamma$--ray emission together with radio to X-ray emission from Mrk 501 during non-flaring activity has been characterized using multi-wavelength campaign between March and
May 2008 \cite{Aleksic2015}. In this study, Mrk 501 is found in low state of activity with significant flux variations in different energy bands and a trend of increasing variability with energy has also been observed. Abdo et al. (2011) have used extensive radio to TeV data set from the multiwavelength campaign of Mrk 501 between March 15, 2009 and August 1, 2009 and have provided the most detailed study of the source during its relatively low activity state \cite{abdo+11+501}.
In this study the authors have presented a detailed description of $\gamma$--ray activity measured by \emph{Fermi}-LAT during its first 16 months
of operation and they have observed only mild $\gamma$--ray flux variations but a remarkable spectral variability. 

\subsection{Multi-wavelength Variability} \label{vaibility}
\begin{figure}[t]
    \centering
    \includegraphics[width=1.0\textwidth]{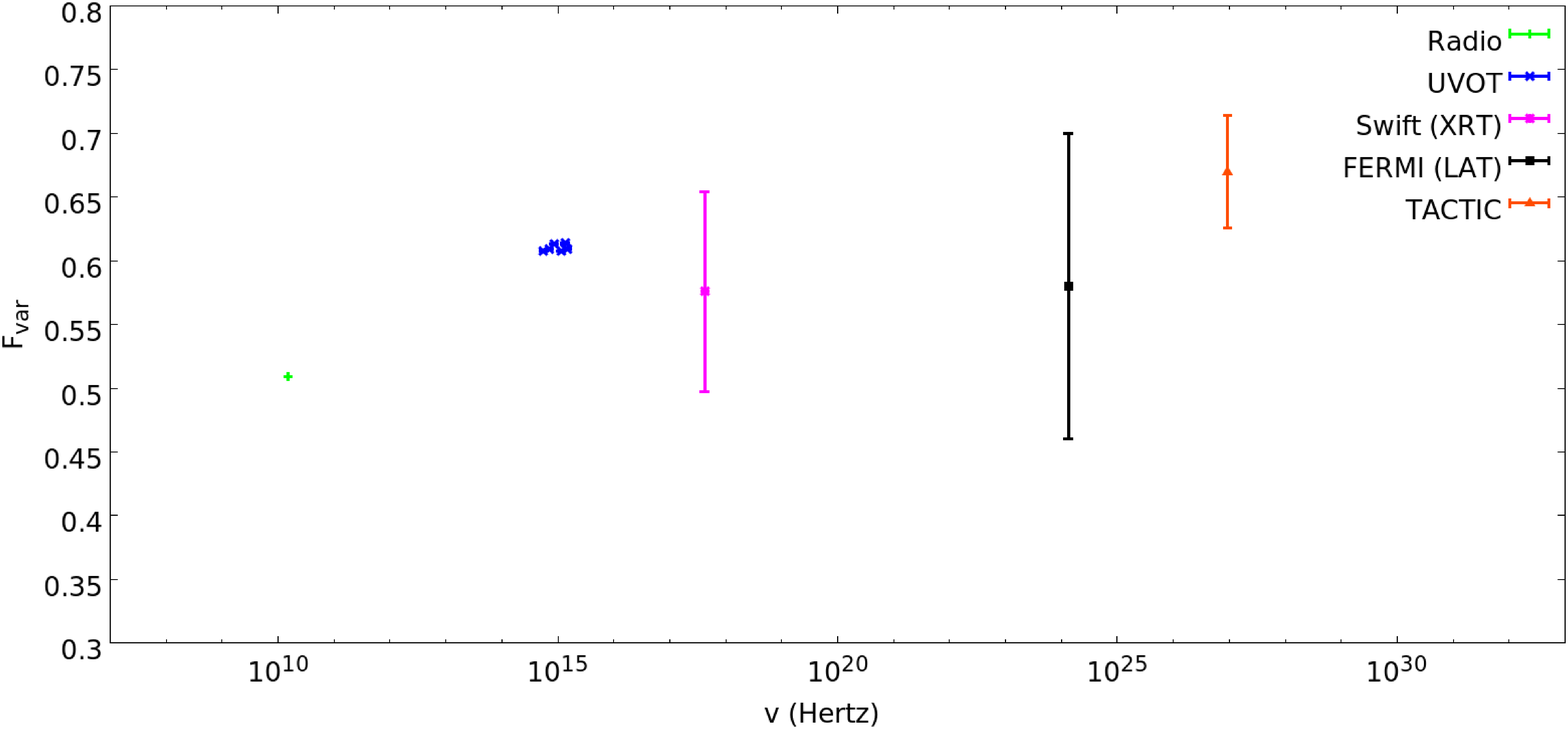}
    \caption{Energy dependence of fractional variability amplitude in the multi-wavelength observation of Mrk 501 during April - May, 2012.} 
    \label{fig:variability}
\end{figure}

The multi-wavelength emission from blazars is usually observed to be variable in almost all energy bands from radio to VHE $\gamma$-rays. 
To quantify the amplitude of variability in various energy bands, we have computed the fractional variability amplitude using one day 
binned light curves in each energy band. The normalized fractional variability amplitude (F$_{var}$) is defined as \cite{Edelson2002},
\begin{equation}
	F_{var}=\frac{\left[S^{2}-\langle \sigma_{err}^{2}\rangle \right]^{\frac{1}{2}}}{\langle F\rangle} 
\end{equation}
where $\langle F\rangle$ is mean photon flux, \textit{S} is standard deviation  and $\sigma_{err}^{2}$ is  mean square error of N-flux points 
in the light curve. The error in fractional variability amplitude is then given by, 
\begin{equation}
       \Delta 	F_{var}=\frac{1}{F_{var}} \sqrt{\frac{1}{2N}} \frac{\langle \sigma_{err}^{2} \rangle}{{\langle F \rangle}^2}
\end{equation}
The $F_{var}$ as a function of mean observational energy of various energy bands is depicted in Figure~\ref{fig:variability}.
 From the figure, it is evident that the multi-wavelength emission from Mrk 501 can be characterized by an average variability 
amplitude of $\sim$ 60$\%$ from radio to TeV $\gamma$--rays. However, due to large uncertainty in the amplitude values it is not possible to quantify 
the exact physical mechanism and energy dependence of F$_{var}$ during the emission from the source. The similar variability amplitudes 
in different energy bands favour synchrotron and synchrotron self Compton process for multi-wavelength emission from the blazar Mrk 501.
\begin{table}[ht]
\caption{Details of energy dependence of fractional and amplitude variability in the multi-wavelength observation of Mrk 501 during April - May, 2012.}
\begin{center}
\begin{tabular}{ c c c c }
  \hline
 Instrument & Energy Band  & F$_{var}$ &  F$_{max}$/F$_{min}$  \\
    \hline
    
            TACTIC                      &      0.850 - 17 TeV        &    0.67 $\pm$ 0.04        &  3.42     \\
            $\it{Fermi}$-LAT            &      0.1 - 300 GeV         &    0.58 $\pm$ 0.12        &  2.60    \\
            $\it{Swift}$-XRT            &      0.3 - 10 KeV          &    0.58 $\pm$ 0.08        &  1.80    \\
			$\it{Swift}$-UVOT           &      2.29 - 6.11 eV        &    0.61 $\pm$ 0.01        &  1.17     \\
            OVRO                        &      15 GHz              &    0.51 $\pm$ 0.01        &  1.05      \\ 
  \hline
\end{tabular}
\end{center}
\label{tab:table_mrk501_variability}

\end{table}

The fractional variability (F$_{var}$) and ratio of maximum and minimum flux (F$_{max}$/F$_{min}$) obtained in the present study are given in Table~\ref{tab:table_mrk501_variability}.

\subsection{Spectral Energy Distribution (SED)} \label{sed}
The near simultaneous multi-wavelength data of Mrk 501 recorded during the TACTIC observations make it possible to 
investigate the average SED of the source. In the present work, we use conventional homogeneous 
single zone synchrotron and synchrotron self Compton (SSC) stationary model \cite{Sunder2012} to describe the SED of the 
blazar Mrk 501. In this model, broadband SED is interpreted as synchrotron and inverse Compton emission processes
respectively from a non-thermal distribution of relativistic electrons confined within a spherical blob moving towards the observer inside the jet. 
The blob is characterized with radius (R), bulk Lorentz factor ($\Gamma$) and tangled magnetic field (B). The energy distribution 
of relativistic electrons in the blob is assumed to be broken power law with indices $n1$ and $n2$ before and after the break energy 
respectively. The respective minimum, maximum and break energies of the electrons correspond to the Lorentz factors $\gamma_{min}$, 
$\gamma_{max}$ and $\gamma_{br}$ respectively. The blob radius is constrained by the variability time scale ($t_{var}$) estimated from 
the light curve using the relation $R \approx \frac{c \delta t_{var}}{(1+z)}$ where $\delta =[\Gamma(1-\beta cos \theta)]^{-1}$, is the Doppler 
factor with $\beta$ being the dimensionless bulk velocity and $\theta$ is viewing angle between the jet axis and line of sight of the observer. 
For blazars, jet is aligned close to the line of sight, we approximate $\delta=\Gamma$ corresponding to $\theta=cos^{-1} 
(\beta)$. The model has typically seven main parameters : R, B, $\delta$, n1, n2, $\gamma_{min}$ and $\gamma_{max}$ and demands a good quality 
multi-wavelength data to constrain them. 
\par
\begin{figure}
    \centering
    \includegraphics[width=0.8\textwidth]{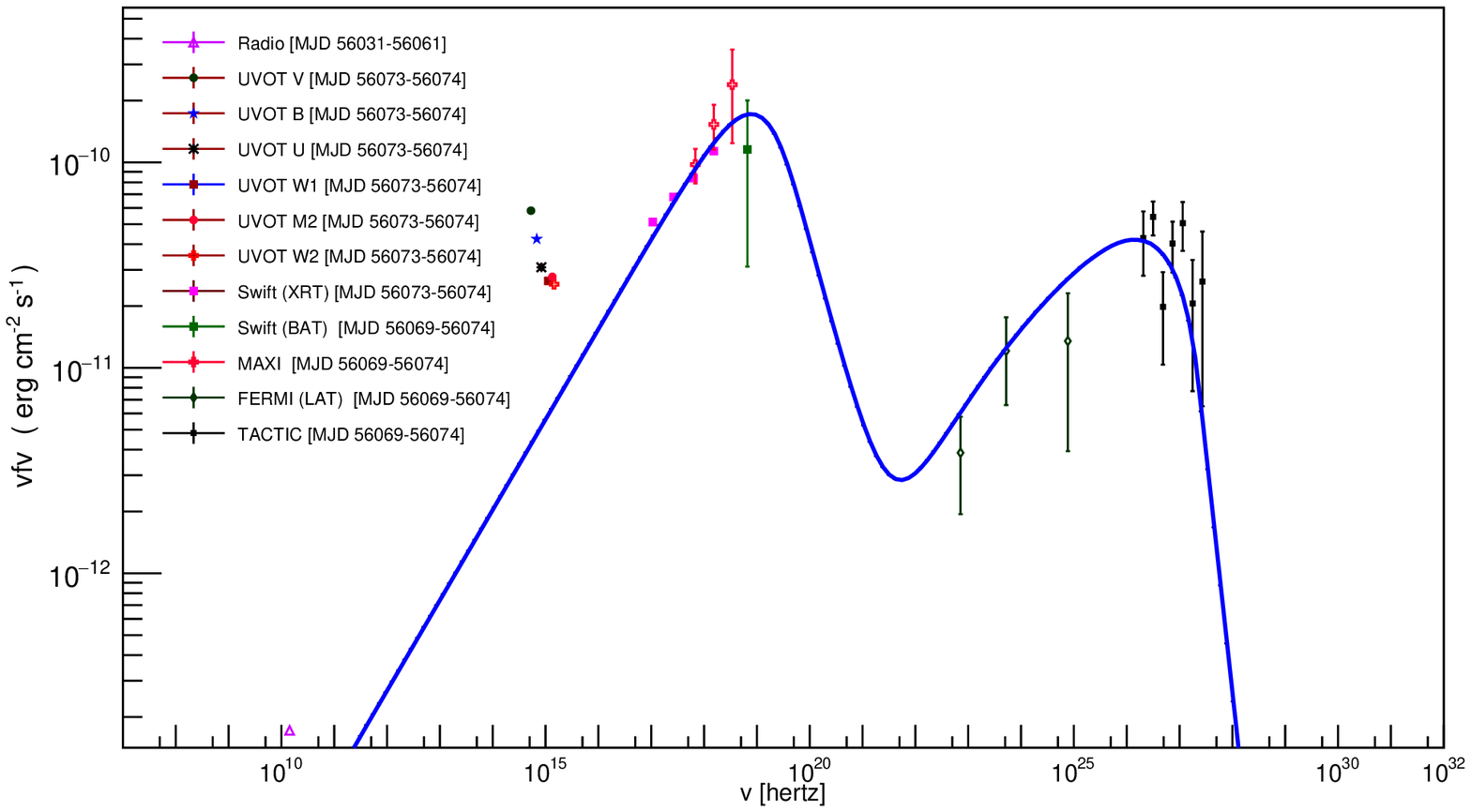}
    \caption{Spectral energy distribution for Mrk 501 during 22-27 May, 2012. The legend report instrument and duration in MJD. The TeV data from TACTIC have been corrected for the absorption due to extragalactic background light using the model reported in Franceschini model ~\cite{franceschini2008}. } 
    
    \label{fig:spectrum_mrk501_sed}
\end{figure}

We use the multi-wavelength data set discussed in previous subsections to construct the SED of Mrk 501 under the framework of the model described 
above. For VHE $\gamma$--rays, we make use of data from TACTIC observations corrected for EBL absorption due to pair production using the 
Franceschini model \cite{franceschini2008}. The \emph{Fermi}-LAT spectrum during the TACTIC observations covers the high energy 
$\gamma$--ray band in the energy range 0.1-300 GeV. The X-rays spectra from \emph{Swift}-XRT, BAT and MAXI have been used in the range from 
0.1-10 keV, 15-50 keV and 10-20 keV respectively. The UV-optical data from UVOT telescope and 15 GHz radio data from OVRO observatory are also 
used for constructing the multi-wavelength SED. We reproduce the SED of Mrk 501 adjusting the parameters of the model to obtain eyeball fit 
which agrees with the multi-wavelength data as is clear in the Figure~\ref{fig:spectrum_mrk501_sed}. From the figure it is evident that the SSC model discussed above reasonably describes the SED of Mrk 501 observed during 22 - 27 May, 2012 except at optical/UV energies. The excess in the optical/UV flux reflects presence of significant host galaxy contribution and hence we omit this data while reproducing the jet non-thermal emission \citep{abdo+11+501}. The obtained model parameters of the source are listed in Table~\ref{tab:sed_parameter}.

\begin{table}[ht]
\caption{The model parameters of Mrk 501 using one zone homogeneous SSC model.}
\begin{center}
  \begin{tabular}{ c c }
    \hline
       Parameter & Value \\
      \hline
	         $\delta$  & 14.2 \\
	          B  & 0.12 G \\
	          R  &  6.1$\times$10$^{15}$cm \\	          
  	          n1  &  2.12 \\	  
  	          n2  &  4.90 \\	  
  	          $\gamma_{min}$ & 50\\	  
   	          $\gamma_{max}$ & 1.0$\times 10^{8}$\\	  
   	          u$_{e}$  & 3.1$ \times 10^{-2} erg/cm^{-3}$\\	    	          
  \hline         
   \end{tabular}
   \end{center}
   \label{tab:sed_parameter}
\end{table}  
\section{Discussion and Conclusions} \label{discuss}
In this paper,  we have studied  a relatively high flux state of blazar Mrk 501 using  TACTIC gamma-ray telescope and     
nearly  simultaneous observations in HE with $\it{Fermi}$-LAT,  X-ray with $\it{Swift}$-XRT and MAXI, UV-optical with \emph{Swift}-UVOT 
and radio with OVRO. Detailed analysis of the TACTIC data recorded during 22 - 27 May, 2012  shows an evidence
of a statistically significant  excess of VHE gamma-ray like events  $375 \pm 47$  with a statistical significance of 8.05$\sigma$ above 850 GeV.  
During this period time-averaged differential energy spectrum of the source in the energy range 850 GeV - 17.24 TeV fits 
well with the power law function  with $f_{0}=  (2.27 \pm 0.38) \times 10 ^{-11} $ photons cm$^{-2}$ s$^{-1}$ TeV$^{-1}$ and $\Gamma=2.57  \pm  0.15$ with a $\chi^{2}$/ndf  value of $\sim$ 9.027/6.  
The blazar Mrk 501 is observed to be variable during TACTIC observation from April-May, 2012 in all energy bands with the fractional variability amplitude in the range $ 50-70\%$. The variability amplitude in radio is about 50$\%$ and 61$\%$ in optical observed with \emph{Swift}-UVOT, while that in the soft 
X-rays is about 58$\%$ in the band 0.3-10 keV observed with \emph{Swift}-XRT.  The variability 
amplitude in MeV-GeV $\gamma$--rays observed with \emph{Fermi}-LAT  is $\sim$58$\%$ , which is less than that in VHE $\gamma$--rays $\sim$67$\%$. 
A trend of increasing variability with energy is observed and is also found to be consistent with the result reported by \cite{Aleksic2015}.
The variability of the source in different energy bands indicates the jet origin of multi-wavelength emission. However, it is not 
possible to identify the exact energy dependence of $F_{var}$ due to large error bars as shown in Figure~\ref{fig:variability}.

\par
The SED of Mrk 501 is well reproduced by a stationary single zone SSC model with a broken power law 
electron energy distribution. We also tried to model the SED using an electron energy distribution with two breaks, as reported in Abdo et al. (2011) \citep{abdo+11+501}, but the available data are not sufficient to disentangle between a model with a single or double broken power law distribution of the electrons. The difference between particle spectral indices $n1$ and $n2$ is greater than one and hence production of 
such spectrum is not feasible by considering only synchrotron and inverse Compton losses. Alternatively, consideration of multiple acceleration processes can develop a broken power-law particle distributions with indices differing more than one \cite{pope1994,sunder2008}. One such scenario can be reacceleration of turbulent accelerated particle at shock fronts. 
The detailed modelling of such scenarios demand high quality multi-wavelength data, which is beyond the scope of this paper. Based on the near simultaneous multi-wavelength data available during the TACTIC observations of Mrk 501, we conclude that one zone SSC model is sufficient to reconstruct the average SED of the source. 
Furniss et al. (2015) have also used single zone homogeneous SSC model to reproduce the broad-band SED of Mrk 501 using multiwavelength observations during April-August 2013  \cite{Furniss2015}. In another study by Aleksic et al. (2015) the authors have studied the broad-band SED of Mrk 501 in two different emission states and have concluded that the higher state can be described by an increase in the electron number density \cite{Aleksic2015}.
Multi-zone SSC model has been proposed for reproducing the broad band SED of Mrk 501 using the multi-wavelength observations during 2011  \cite{Shukla2015}. The study suggests that the emission from Mrk 501 is 
a complex superposition of multiple emission zones and indicates the existence of two separate components in the spectrum for 
low and high energy $\gamma$--rays. In another multiwavelength study of the source  \cite{Furniss2015} the authors have used single zone equilibrium SSC model to reproduce the five simultaneous multi-wavelength SEDs of the source. They find that SSC model with a power law 
distribution of electrons can reproduce the observed broad band states through a decrease in the magnetic field coinciding with 
an increase in the luminosity and hardness of the relativistic leptons responsible for high energy emission. 
Further in paper  \cite{Bartoli2012} authors have used a time-dependent SSC model with electron energy distribution 
parameterized by a single power law with an exponential cutoff at its high energy end to describe the multi-wavelength SED of 
Mrk 501 observed in flaring state during October-November 2011. The average SED of the source for steady emission 
is also well described by this one zone SSC model, but the detection of VHE $\gamma$--rays above 8 TeV during the flare challenges 
this model due to hardness of the spectra. We have planned to observe the source using TACTIC telescope in future to understand 
the source emission models in detail with concurrent multiwavelength observations. 	

\section{Acknowledgements} 
Authors thank the  anonymous reviewer for his/her valuable comments which have helped in improving the manuscript.
We would like to acknowledge the excellent team work of our colleagues in the Division for their contribution to the simulation, observation, data analysis and instrumentation aspects of TACTIC at Mt. Abu observatory.
We acknowledge the use of public data obtained through \textit{Fermi} Science Support Center (FSSC) provided by NASA. This work made use of data supplied by the UK Swift Science Data Centre at the University of Leicester. This research has made use of the MAXI data, provided by RIKEN, JAXA and the MAXI team. Data from the Steward Observatory spectropolarimetric monitoring project were used. This program is supported by  \emph{Fermi} Guest Investigator grants NNX08AW56G, NNX09AU10G and NNX12AO93G.
Radio data at 15 Ghz is used from
OVRO 40 M Telescope and this \emph{Fermi} blazar monitoring program is supported by NASA under award NNX08AW31G,
and by the NSF under 0808050. 





\end{document}